\documentclass[11pt]{revtex4-1}

\usepackage{amsmath}
\usepackage{graphicx}
\usepackage{xcolor}
\usepackage{courier}
\usepackage{hyperref}

\begin{document}
    \title{Highly accurate potential energy curves for the hydrogen molecule ion}

    \author{Francisco M. Fern\'{a}ndez}
    \author{Javier Garcia}
    \email{jgarcia@fisica.unlp.edu.ar}
    \affiliation{INIFTA, DQT, Diagonal 113 y 64 S/N 1900 La Plata, Argentina}

    \begin{abstract}
        Potential energy surfaces of the hydrogen molecular ion H$_2^+$ in the Born-Oppenheimer approximation are computed by means of the Riccati-Pad\'e method (RPM).
        The convergence properties of the method are analyzed for different states.
        The equilibrium internuclear distance, as well as the corresponding electronic plus nuclear energy, and the associated separation constants, are computed to 40 digits of accuracy for several bound states.
        For the ground state the same parameters are computed with more than 100 digits of accuracy.
        Additional benchmark values of the electronic energy at different internuclear distances are given for several additional states.
        The software implementation of the RPM is given under a free software license.
        The results obtained in the present work are the most accurate available so far, and further additional benchmarks are made available through the software provided.
    \end{abstract}

    \maketitle

    \section{Introduction}

    The hydrogen-ion molecule H$_2^+$ is the simplest molecule that exists in Nature, and, after applying the Born-Oppenheimer approximation, the solution of the Schr\"odinger equation becomes one of the simplest non-trivial problems in quantum mechanics.
    The Schr\"odinger equation is separable into two one-dimensional equations that depend on the electronic energy, the internuclear distance, and a separation constant, and the relative ease of their solution led to excellent analyses of the H$_2^+$ spectrum as early as the 1950s \cite{Bates_1953, Peek_1965, Bates_1968, Beckel_1970, Madsen_1970}.
    These efforts are nothing short of remarkable, given the time period they were performed at, but due to the obvious limitations imposed by the computing power available at the time, the tabulated spectra were computed using double precision, meaning that they contain at most 15 significant digits.

    In general, more accurate computations of the eigenspectra of quantum-mechanical problems are of interest for testing other computational methods (see, for example, the recent discussion in Refs. \cite{Okun_2020} and \cite{Turbiner_2021}).
    The H$_2^+$ system is of particular interest since, as it involves a real molecule, it is useful not only to provide benchmark values for testing numerical methods \cite{Braun_2012} in general, but also for and quantum-chemical methods, such as those based on perturbation theory \cite{Chipman_1973, Jeziorski_1978, Chalasinaski_1980}, or the variational theorem \cite{Yamamoto_2017,Sarwono_2020}. 
    It has also been used to validate potential energy surfaces (PESs) \cite{Xie_2005, Xie_2014}, which are of the utmost importance in the computation of condensed matter properties \cite{Szalewicz:14a, Metz_2016}.
    The computed spectra and PESs of H$_2^+$ have also been used to model covalent bonding \cite{Schmidt_2014}, and in experimental observations of molecular properties \cite{Beyer_2016, Beyer_2016a, Schiller_2017}.
    Both the lower- and higher-lying states of the H$_2^+$ have been recently computed using semiclassical approximations \cite{Olivares_Pil_n_2016, Khmara_2018, Price_2018}, and variational approaches\cite{de_Oliveira_Batael_2020}.
    The solutions of the H$_2^+$ molecule ion problem have been also analyzed in terms of hypergeometric functions \cite{Figueiredo_1993}, the Heun confluent functions \cite{Figueiredo_2007, Boyack_2011}, and Coulomb Sturmians \cite{Kereselidze_2015, Kereselidze_2015a}.
    Series solutions have also been given for the spectrum of H$_2^+$ \cite{Scott_2006}.
    
    Upon thorough inspection of the literature surveyed here, it is found that all the tabulated spectra contain at most 15 significant digits, which is a product of the limitation of floating point calculations.
    Some of the references cited here describe methods that can be programmed using computer albebra software (CAS), but none of them provide software that is independent of them.
    The only referenced software that meets this requirement is the program ODKIL\cite{Hadinger_1989}, which is written in FORTRAN and only allows for floating point computations.

    The Riccati-Pad\'e method (RPM) \cite{Fernandez_1989,Fernandez_1989a,Fernandez_1989b,Fernandez_1989c, Fernandez_1992,Fernandez_1993,Fernandez_1995} is a very straightforward method to solve the Schr\"odinger equation and related eigenvalue problems that consists in writing a Riccati equation for the derivative of the logarithm of the wave function and using increasingly large Pad\'e approximants to represent it, in such a way that the Taylor expansion of both the exact solution and the Pad\'e approximant coincides up to one extra coefficient than the definition of Pad\'e approximant accounts for.
    This leads to a quantization condition that involves finding the root of the determinant of a Hankel matrix built with the expansion coefficients.
    The RPM can be programmed with very little effort using a CAS, and it has a very fast rate of convergence to the eigenvalues of the Schr\"odinger equation and related problems \cite{Fernndez2016, Fern_ndez_2017, Fernndez2018}.
    In the present work we provide an efficient implementation of the RPM to solve the Schr\"odinger equation for the H$_2^+$ that is independent of CAS.
    We use our implementation to thoroughly test the convergence properties of the RPM, and we compute the spectrum of 69 different states to 8 -- 25 digits of accuracy.
    We select a few of those states and perform computations with 60 -- 100 digits of accuracy; these results may be used as benchmarks for testing other methods.
    Finally, we analyze the spectra we computed and identify the bound states; we compute the internuclear distance, electronic energy, and separation constant to $\sim$ 40 digits of accuracy.
    For the ground state, this computation is refined to produce 160 significant digits.
    We briefly describe the software that allowed such computations, which is distributed under a free software license. 

    \section{The method}

    The RPM has been thoroughly described in previous works \cite{Fernandez_1989, Fernandez_1989a, Fernandez_1989b, Fernandez_1989c, Fernandez_1992, Fernandez_1993, Fernandez_1995, Fernandez_1996, Fern_ndez_2016, Fern_ndez_2017, Fernndez2018}, but for the sake of completeness we briefly recall its main features here.
    We also revisit the main generalities of the Schr\"odinger equation for H$_2^+$, and detail the application of the RPM to solve it.

    \subsection*{The RPM for coupled equations}

    The Schr\"odinger equation and related problems can usually be written in the following manner,
    \begin{equation}
        L^{\prime\prime}(x) + P(x) L^\prime(x) + Q(x) L(x) = 0.
        \label{eq:schrodinger_generic}
    \end{equation}
    where $P(x)$ and $Q(x)$ are arbitrary functions that admit expansions in powers of $x$, $P(x) = \sum_{k = -1}^\infty p_k x^k$, and $Q(x) = \sum_{k = -2}^\infty q_k x^k$ and depend on one or more parameters $\lambda_i$.
    In the case of the one-dimensional Schr\"odinger equation, for example, $P(x) = 0$, and $Q(x) = 2E - V(x)$.
    If $L(x) \sim x^s$ when $x \rightarrow 0$, with $s$ being an integer number, then the the regularized logarithmic derivative of $L$, i.e., the function
    \begin{equation}
        f(x) = \frac{s}{x} - \frac{L^\prime(x)}{L(x)},
        \label{eq:f}
    \end{equation}
    can be expanded in a Taylor series around $x = 0$, i.e., $f(x) = \sum f_j x^j$,  and satisfies the following Riccati equation,
    \begin{equation}
        f^\prime(x) + \left [ \frac{2s}{x} + P(x) \right ] f(x) - f^2(x) - \frac{s}{x} P(x) - Q(x) -\frac{s(s-1)}{x} = 0.
        \label{eq:riccati_generic}
    \end{equation}
    A recurrence relation can be found that relates each $f_j$ with the preceding $f_{0}, \ldots, f_{j-1}$, and these also depend on the same parameters as $Q(x)$ and $P(x)$.

    We now consider an $[M/N]$ Pad\'e approximant to $f(x)$, i.e., a quotient of polynomials of degrees $M$ and $N$, 
    \begin{equation}
        [M/N](x) = \frac{ \sum_{j=0}^M a_j x^j}{1 + \sum_{k=1}^N b_k x^k} ,
    \end{equation}
    such that the series expansion of both $f(x)$ and $[M/N](x)$ coincide up to order $M+N+1$.
    To each function $f(x)$, there is a unique $[M/N]$ Pad\'e approximant, and there is a set of $M+N$ linear equations that relate the coefficients $a_j$, and $b_j$ with the $f_j$ coefficients.
    The RPM consists in choosing the parameters on which $P$ and $Q$ depend in such a way that $[M/N]$ matches an extra coefficient $f_j$, i.e., $[M/N](x) - f(x) = O(x^{M+N+2})$.
    This adds an extra equation to the set of $M+N$ ones, and it is straightforward to show that this set has a non-trivial solution if
    \begin{equation}
            H_D^d(\lambda_1, \ldots, \lambda_n) = \left | 
            \begin{array}{cccc}
            f_{d+1} & f_{d+2} & \ldots & f_{d+D} \\ 
            f_{d+2} & f_{d+3} & \ldots & f_{d+D+1} \\ 
            \vdots & \vdots & \ddots & \vdots \\ 
            f_{d+D} & f_{d+D+1} & \ldots & f_{2D+d-1}
            \end{array} \right | = 0,
            \label{eq:hankdet}
    \end{equation}
    where we have defined $D = N+1$, and $d = M - N$.

    Eq. \eqref{eq:hankdet} provides a quantization condition for one of the parameters $\lambda_1, \ldots, \lambda_n$ and allows to obtain it in terms of the others.
    For example, in the case of the one-dimensional Schr\"odinger equation, if the Hamiltonian depends on one parameter, then one can obtain the energy in terms of it, or the values of said parameter for which the energy adopts a particular value, as in the case of the critical parameters\cite{Fernandez_2013}.
    In the case of coupled equations (which is of concern in the present work), one should have as many coupled equations as unknown parameters, and Eq. \eqref{eq:hankdet} should be solved simultaneously for each of the equations.
    The quantization condition \eqref{eq:hankdet} is known to yield both the bound states and resonances of many problems in Quantum Mechanics, and it is believed it does so by sending a movable pole at complex infinity \cite{Abbasbandy:11}.
    The singularity can be moved around any path in the complex plane, meaning that Eq. \eqref{eq:hankdet} also gives solution to both kinds of problems without specifying the boundary conditions \cite{Fernandez_1996, Fernndez2016}.

    \subsection*{Application to the eigenenergies of the H$_2^+$ molecule-ion}

    Within the Born-Oppenheimer approximation, the electronic Hamiltonian for the H$_2^+$ molecule is,
    \begin{equation}
        H_{\rm mol} = -\frac{1}{2} \nabla^2 - \frac{1}{r_1} - \frac{1}{r_2},
        \label{eq:H}
    \end{equation}
    where $\nabla$ involves differentiation with respect to the electron coordinates, and $r_{1, 2}$ are the absolute distances between the electron and each of the nuclei.
    Here atomic units are being used. 
    The Schr\"odinger equation defined by Eq. \eqref{eq:H} can be transformed into a set of separable equations by using the spheroidal coordinates $\lambda, \mu$ and $\phi$, where $\lambda$ and $\mu$ are defined as 
    \begin{align}
        \lambda &= \frac{r_1 + r_2}{R}, 1 \leq \lambda < \infty, \\
        \mu &= \frac{r_1 - r_2}{R}, -1 \leq \mu \leq 1, 
    \end{align}
    and $0 \leq \phi \leq 2\pi$ is the angle of rotation of the electron around the internuclear axis.
    $R$ is the internuclear separation, also in atomic units.
    The electronic wavefunction can be written as a product $\psi(\lambda, \mu, \phi) = L(\lambda) M(\mu) \Phi(\phi)$, and satisfies the following equations:
    \begin{align}
        & \Phi(\phi) = \frac{1}{\sqrt{2\pi}} e^{im\phi}, \quad m = 0, \pm 1, \pm 2, \ldots , \\
        & \frac{d}{d\lambda} \left [ (\lambda^2 - 1) \frac{d L(\lambda)}{d \lambda} \right ] + \left [ -\frac{m^2}{\lambda^2 - 1} - \epsilon \lambda^2 + 2 R \lambda + A \right ] L(\lambda) = 0 \label{eq:lambda}\\
        & \frac{d}{d\mu} \left [ (1 - \mu^2) \right ] + \left [ -\frac{m^2}{1- \mu^2} + \epsilon \mu^2 - A \right ] M(\mu) = 0, \label{eq:mu}
    \end{align}
    where $A$ is the separation constant, $\epsilon = -R^2 E/2$, with $E$ being the electronic energy i.e., the eigenvalue of $H_{\rm mol}$, and $m$ is the quantum number associated with the angular momentum in the direction of the internuclear axis.

    In order to solve Eqs. \eqref{eq:lambda} and \eqref{eq:mu} with the RPM, we first define $x = \lambda -1$, which transforms Eq. \eqref{eq:lambda} into Eq. \eqref{eq:schrodinger_generic}, with
    \begin{equation}
        \begin{aligned}
            P_\lambda(x) &= \frac{2(x+1)}{x(x+2)}, \\
            Q_\lambda(x) &= \frac{2R(x+1)+A-\epsilon(x+1)^2}{x(x+2)} - \frac{m^2}{x^2(x+2)^2}.
        \end{aligned}
    \end{equation}
    The solution $L(x)$ is known to behave at origin as $x^{|m|/2}$; then by setting $s = |m|/2$ in Eq. \eqref{eq:f} we remove the singularities of Eq. \eqref{eq:schrodinger_generic} at origin, and writing $f(x) = \sum_{j=-1}^\infty f_j x^{j+1}$, the following recurrence relation for the coefficients $f_j$ is obtained,
    \begin{equation}
        \begin{aligned}
            f_{-1} &= \frac{2s^2 + s - \epsilon + 2R + A}{2(2s+1)}, \\
            f_j &= \frac{1}{n+2s+p^\lambda_{-1}+1} \left [ 
        \sum_{k=0}^j \left ( f_{k-1}f_{j-k-1} - 
        p^\lambda_k f_{j-k-1} + s p^\lambda_{k+1} \right ) + q^\lambda_j 
        \right ], j = 0, 1, \ldots\,,
        \end{aligned}
    \end{equation}
    where $p_j^\lambda$ and $q_k^\lambda$ are the expansion coefficients for $P_\lambda(x)$ and $Q_\lambda(x)$.

    Similarly, by setting $\mu = x$, Eq. \eqref{eq:mu} transforms into Eq. \eqref{eq:schrodinger_generic}, with
    \begin{equation}
        \begin{aligned}
            P_\mu(y) &= -2x / (1-x^2), \\
            Q_\mu(y) &= \frac{\epsilon x^2 - A}{1-x^2} - \frac{m^2}{(1-x^2)^2}.
        \end{aligned}
    \end{equation}

    Now we define
    \begin{equation}
        g(x) = \frac{t}{x} - \frac{M^\prime(x)}{M(x)}.
    \end{equation}

    Since $P_\mu(x)$ is odd and $Q_\mu(x)$ is even, $M(x)$ has defined parity.
    Therefore, $P_\mu(x) = \sum_{j = 0}^\infty p_j^\mu x^{2j+1}$, $Q_\mu(x) = \sum_{j=0}^\infty q^\mu_j x^{2j}$, and $g(x) = \sum_{j=0}^\infty g_j x^{2j+t}$, where $t = 0$ for even states and $1$ for odd states, and the following recurrence relation is obtained:
    \begin{equation}
        g_j = \frac{1}{2j + 2t + 1} \left [ 
            \sum_{k=0}^j \left ( 
                f_k f_{j-k-1} - p^\mu_k f_{j-k-1} \right )
                + t p^\mu_j + q^\mu_j \right ], \quad j = 0, 1, 2, \ldots\,.
    \end{equation}

    The RPM quantization condition \eqref{eq:hankdet} becomes
    \begin{equation}
        H_{\mu, D}^d(\epsilon, A) = H_{\lambda, D}^d(\epsilon, A, R) = 0.
        \label{eq:hankdet_roots}
    \end{equation}
    There are three unknowns and only two equations; consequently, equations (10) and (11) yield two parameters in terms of the third one.
    For benchmarking purposes, it is customary to solve for $\epsilon$ and $A$ in terms of $R$.

    \subsection*{Computation of the Hankel determinants and their roots}

    Any computer algebra system (CAS) contains subroutines to compute the determinant of a matrix. 
    These subroutines are usually good enough for general purposes, but they don't exploit the recurrence relation obeyed by the Hankel matrices that provides a much more efficient way to compute them,
    \begin{equation}
        H_D^d = \frac{ 
            H_{D-1}^d H_{D-1}^{d+2} - (H_{D-1}^{d+1})^2
        } {
            H_{D-2}^{d+2}
            },
            \label{eq:hankdet_recurr}
    \end{equation}
    where $H_1^d = f_{d+1}$, and we define $H_0^d = 1$. 
    To use Eq. \eqref{eq:hankdet_recurr}, we start by computing coefficients $f_{d+1}, \ldots, f_{2D+d-1}$.
    Then, we use them to compute $H_2^{d}, \ldots, H_2^{2D+d-4}$; these are used to compute $H_3^d, \ldots, H_3^{2D+d-6}$, and so on, until we reach $H_D^d$.
    Since only determinants of order $D-1$ and $D-2$ are needed to compute determinants of order $D$, for each step we only need to store two rows of determinants in memory.
    As an example, Figure \ref{fig:sequence} shows the sequence of determinants required to compute $H_5^2$.

    \begin{figure}
        \begin{equation*}
            \begin{array}{ccccccccc}
                f_3   & f_4   & f_5   & f_6   & f_7   & f_8   & f_9   & f_{10}   & f_{11}   \\
H_2^2 & H_2^3 & H_2^4 & H_2^5 & H_2^6 & H_2^7 & H_2^8 &       &       \\
H_3^2 & H_3^3 & \mathbf{H_3^4} & H_3^5 & H_3^6 &       &       &       &       \\
                \mathbf{H_4^2} & \mathbf{H_4^3} & \mathbf{H_4^4} &       &       &       &       \\
                \mathbf{H_5^2} &       &       &       &       &       &       \\
            \end{array}
        \end{equation*}
        \caption{Sequence of Hankel determinants needed to compute $H_5^2$. The subset of determinants required for the last step are marked in bold. Only two consecutive rows need to be stored in memory at the same time. \label{fig:sequence}}
    \end{figure}

    Now we discuss the issue of solving Eqs. \eqref{eq:hankdet_roots}, i.e., finding the roots of the Hankel determinants.
    To that end, we resort to the well-known Newton-Raphson (NR) method, using numerical differentiation in the form of the symmetric difference quotient.
    One difficulty that may arise when finding the roots of the Hankel determinants is that $H_D^d$ presents multiple roots that approach to the eigenvalues of the problem \cite{Fernandez_1989c, Fernandez_1995}.
    The reason for this is hinted at Eq. \eqref{eq:hankdet_recurr}, by realizing that one solution for $H_D^d = 0$ is $H_{D-1}^dH_{D-1}^{d+2} = (H_{D-1}^{d+1})^2$, but also if both $H_{D-1}^d$, $H_{D-1}^{d+2}$, and $H_{D-1}^{d+1}$ are close enough to 0 (they should be, since they are themselves the roots of another Hankel determinant), $H_D^d$ may also become 0.
    It is typically observed that different sequences of roots converge towards the desired eigenvalues with different speed; we call the fastest of them the ``main sequence''.
    If the starting values provided for the NR method are not good enough it may converge to roots that are not in the main sequence, and for increasing $D$, the sequence of roots of $H_D^d$ may converge more slowly towards the desired eigenvalue.
    Besides, it is a well-known fact that the convergence of the NR method deteriorates when dealing with multiple roots.
    Despite these issues, we have been able to successfully compute the roots of the Hankel determinants for other problems in the past, and for the problem described here in particular, as will be shown below.

    The simultaneous solution of Eqs. \eqref{eq:hankdet_roots} allows for two different approaches.
    One of them is to choose a desired value of $R$, and then solve both equations simultaneously for $\epsilon$ and $A$.
    This implies solving a system of two equations with two unknowns, for which the NR method is very well suited, but each iteration of the method involves the evaluation of four derivatives.
    The other one is to set a value of $\epsilon$, then solve $H_{\mu, D}^d(\epsilon, A)$ to get the corresponding value of $A$, and solve $H_{\lambda, D}^d(\epsilon, A, R)=0$ for the value of $R$. 
    The latter approach has the advantage that it is only required to solve two separate equations, instead of solving them simultaneously, so that it is only necessary to perform one differentiation per iteration for each equation.
    Unfortunately, the utility of the latter method is limited, since for benchmarking purposes it is usually more desirable to compute $\epsilon$ and $A$ values for selected values of $R$.
    In the present work we will use both approaches with different purposes, as discussed in the following sections.

    The software developed to compute the roots of the Hankel determinants for the $H_2^+$ molecule is made available under a free software license; its location is disclosed in the Supplementary Material section below.

    \section{Results}

    In this section we analyze the rate at which the roots of Eqs. \eqref{eq:hankdet_roots} converge towards the values of the energy and separation constant.
    We then compute the dissociation curves for 69 states; from those curves we identify the bound states and compute the equilibrium internuclear distance, electronic plus nuclear energy, and separation constants to unprecedented accuracy.
    We also provide accurate benchmark values of the electronic energy and separation constant for 21 states at a fixed internuclear distance.

    \subsection*{Convergence of the roots of the Hankel determinants}

\begin{figure}[h]
        \includegraphics[width=\textwidth]{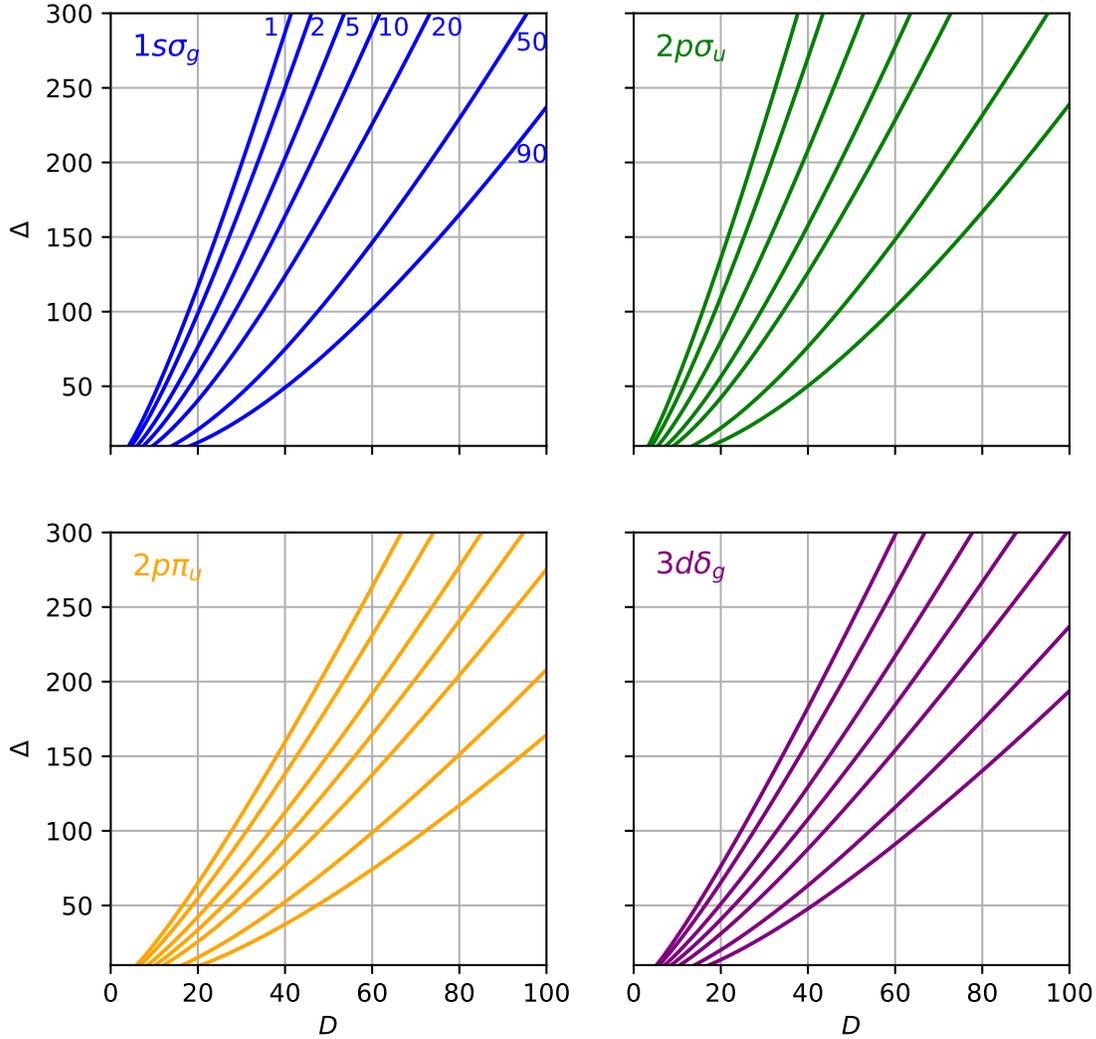}
        \caption{\label{fig:conv-A} Convergence of the roots of $H_{\mu, D}^0 = 0$ towards the correct value of $A$ for different values of $\epsilon$ in different states. 
            Here, $\Delta = -\log_{10} |(A[D,0] - A[D-1,0])/A[D,0]|$ is shown.
        For each graph, the lines from left to right correspond to $\epsilon$ values such that $R = 1, 2, 5, 10, 20, 50$, and $90$ (the values of $R$ are shown next to each line for the state $1s\sigma_g$, and for the other states it follows the same order).}
\end{figure}

\begin{figure}[h]
        \includegraphics[width=\textwidth]{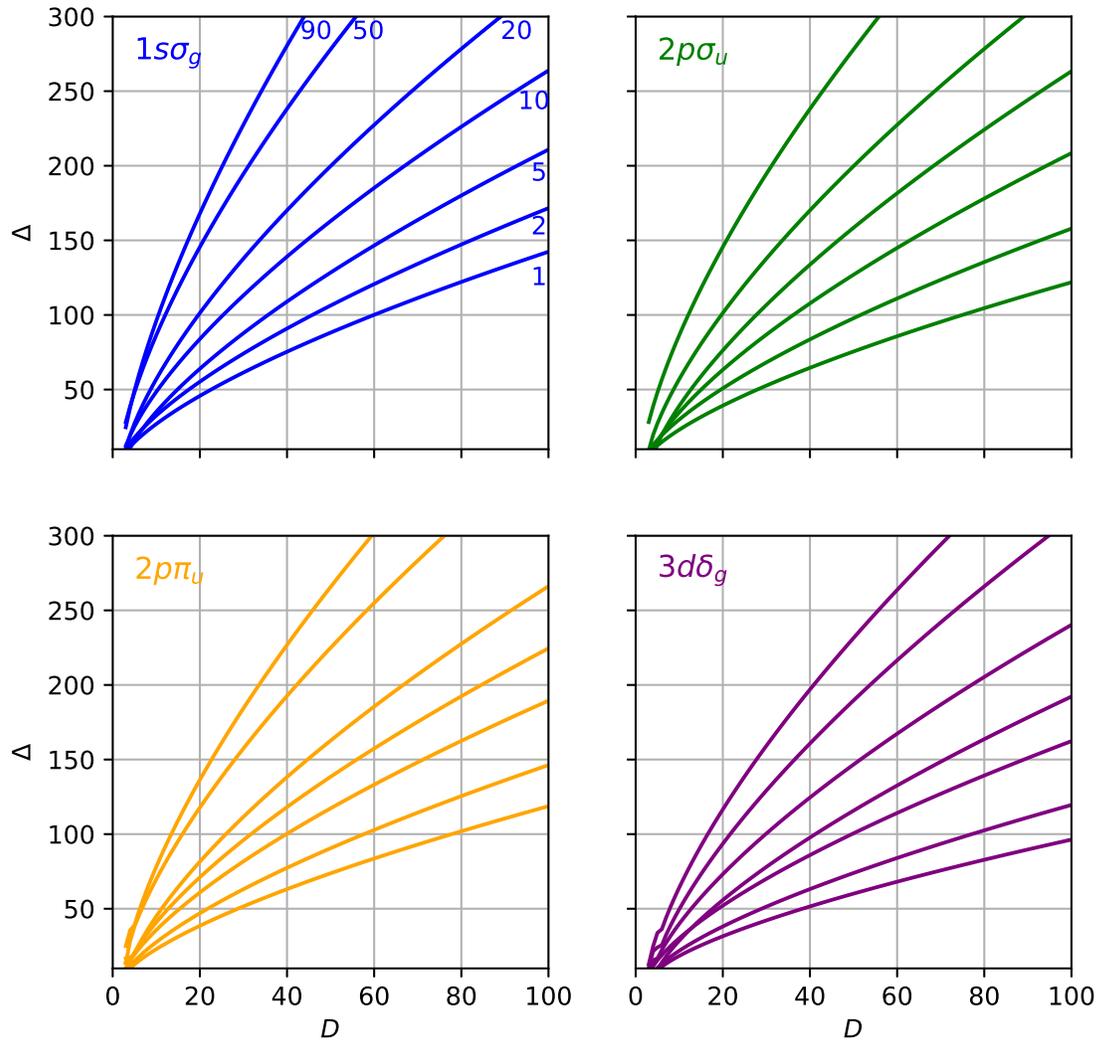}
        \caption{\label{fig:conv-R} Convergence of the roots of $H_{\lambda, D}^0 = 0$ towards the correct value of $R$ for different values of $\epsilon$ in different states. 
        Here, $\Delta = -\log_{10} |(R[D,0] - R[D-1,0])/R[D,0]|$ is shown.
        For each graph, the lines from right to left correspond to $\epsilon$ values such that $R = 1, 2, 5, 10, 20, 50$, and $90$ (the values of $R$ are shown next to each line for the state $1s\sigma_g$, and for the other states it follows the same order).}
\end{figure}

    We first turn our attention to the rate at which the solutions of Eqs. \eqref{eq:hankdet_roots} converge towards the correct values of $A$ and $\epsilon$ for different values of $R$.
    To do so, we first set a value of $R$ ($ =1, 2, 5, 10, 20, 50, $ or $90$), and then we pick an approximate value of the electronic energy $E$ from Ref. \cite{Madsen_1970} (the use of literature values is not strictly required, as discussed in the following section, but facilitates our work).
    We then compute $\epsilon = -R^2E/2$ and solve $H_{D, \mu}^0(\epsilon, A) = 0$ for increasing values of $D$ (using $d = 1, 2$ gives similar results).
    To measure the convergence rate, we plot $\Delta = \log_{10} |(A[D,0] - A[D\raisebox{.75pt}{-}1,0])/A[D,0]|$ against $D$, where $A[D, d]$ is the value of $A$ computed by solving $H_{\lambda, D}^d = 0$. 
    This plot gives a good approximation to the number of significant digits in the result.
    We proceed similarly to measure the convergence rate of the roots of $H_{\lambda, D}^0$ towards the value of $\epsilon$ that corresponds to the value of $R$ chosen at the beginning and the corresponding value of $A$ computed with $H_{\mu, D}^0 = 0$. 
    We repeated this procedure for the states $1s\sigma_g$, $2p\sigma_u$, $2p\pi_u$, and $3d\delta_g$, which are the states with the lowest energy for the sets of quantum numbers $(m, s)$ equal to $(0,0)$, $(0,1)$, $(1,0)$, and $(1,1)$, respectively.
    The plots of $\Delta$ vs $D$ for the $\mu$- and $\lambda$- equations are shown in Figs. \ref{fig:conv-A} and \ref{fig:conv-R}, respectively.

    Figure \ref{fig:conv-A} shows that for the four states analyzed, the smaller the values of $R$ (or $\epsilon$), the greater the rate of convergence.
    Remarkably, the later is faster than exponential, since the plot of the logarithmic difference between iterations is convex.
    An analysis of the plot for $\epsilon$ leads to the opposite conclusion, i.e., convergence rate is slower than exponential, and it increases for larger $R$.
    The result is that for larger (or smaller) values of $R$, solving both of equations \eqref{eq:hankdet_roots} simultaneously can become more difficult, since the solution of one of the equations converges faster than the other one.
    Despite this, for the lower-lying states, convergence speed of both equations is fast enough to obtain accurate results without great computational effort, as shown in the following sections.

    \subsection*{Computation of the spectrum of H$_2^+$}

    As discussed before, for the NR method to converge towards the roots of the Hankel determinants, it is important to provide good initial values. 
    This usually implies computing approximate values using another method or looking them up in literature, but for this particular problem we have designed a very simple algorithm that allows to compute the electronic energy for any state of the H$_2^+$ molecule without resorting to external references.
    To do so, it takes advantage of the fact that when $R = 0$, both nuclei merge, transforming the system into He$^+$, whose Schr\"odinger equation can be solved exactly. 
    We refer to this case as the united-atom (UA) limit.
    The eigenvalue $E$ in the UA limit is equal to $-2/n_u^2$, where $n_u$ is the quantum number (note that $n_u$ is a ``good'' quantum number only in this limit).
    To identify each state for general $R$ and relate it to the UA ones, we follow Ref. \cite{Bates_1968} in using the quantum number $m$, and the ``quantum numbers'' $l$ and $I$.
    Here $l$ is the angular quantum number of the UA solution, and $I = n_u - l$ distinguishes states with the same $l$ but different $n_u$.
    The numbers of nodes of the solutions of equations \eqref{eq:lambda} and \eqref{eq:mu} ($N_\lambda$ and $N_\mu$) are related to $I$ and $l$ according to $N_\lambda = I - 1$, and $N_\mu = |l - m|$.
    As stated before, when computing the Hankel determinants, $s$ must be set to $|m|/2$, and $t$ must be set equal to $0$ if $N_\mu$ is even, and to $1$ if $N_\mu$ is odd.
    The separation constant $A$ in the UA limit is $A = -l(l+1)$ \cite{Scott_2006}.

    To compute the whole energy curve (and its corresponding separation constants), we begin from the UA limit ($R = 0$), computing the corresponding values of $E$ and $A$ using the exact formulas.
    Then, we set $R \leftarrow R + \delta$, and use these values of $E$ and $A$ as starting points for the NR method to solve Eqs. \eqref{eq:hankdet_roots} simultaneously, with $D = D_{\rm min}$, with $D_{\rm min}$ originally set to 5. 
    The result is used as a starting point for $D = D_{\rm min} + 1$, and so on, until $D = D_{\rm min} + 10$.
    This yields results with a varying number of significant digits, depending on the rate at which Eqs. \eqref{eq:hankdet_roots} converge towards the correct values, which can fluctuate significantly, as shown in Figs. \ref{fig:conv-A} and \ref{fig:conv-R}.
    By comparing the amount of digits that coincide between the roots of Eqs. \eqref{eq:hankdet_roots} for two subsequent values of $D$, one can estimate the number of digits of a given approximate result that are correct.
    For a given value of $D$, the Hankel determinant may not be large enough that its roots give an approximation to a particular state.
    For increasing $N_\lambda$ and $N_\mu$, usually larger values of $D$ are required, and if using a smaller value, the NR method converges to an approximation of a different state, or does not converge at all.
    This can be easily accounted for by comparing the converged result with the initial value provided to the NR method. 
    If the relative difference between these values is greater than a set threshold, then the computation is repeated using a larger value of $D_{\rm min}$.
    The new values $E(R + \delta)$ and $A(R + \delta)$ can be used as starting values for $R + 2\delta$, and the procedure is repeated until the desired values of $R$ are covered.
    For a given $R$, too large values of $\delta$ make $E$ and $A$ unsuitable starting points for the NR method to find the Hankel determinant roots for $R + \delta$.
    A much better starting point is provided by extrapolating the three previous values of $E$ and $A$ using a quadratic equation.
    We have included all the computations we performed in the Supplementary Material, with the correct amount of significant digits.
    The algorithm described above is also provided under a free software license in the form of a Python script.

    \subsection*{Accurate benchmark values for selected cases}

    As discussed in the Introduction, it is of interest to have accurate benchmark values available, mainly to use them as a test for other methods.
    Here we compute $E$ and $A$ for some electronic states at particular internuclear distances to a large number of digits.
    We do so by using as starting points the values computed in the previous section, and improve their accuracy by solving Eqs. \eqref{eq:hankdet_roots} for values of $D$ from 2 to 100.
    Whenever we found that a result was accurate to 100 or more digits, we stopped the calculation to save time.
    Also, some of the computations were stopped earlier because the NR method was unable to find the roots after 10000 successive iterations.
    The values provided here may serve as benchmarks for testing other methods, and we would like to reiterate here that results of similar quality can be obtained for different internuclear positions and quantum numbers by using the software provided with the present work.
    For brevity, the complete results are not presented in this article, but they are provided in the Supplementary Material.
    Table \ref{tab:benchmarks} has a summary of all the computed states and the number of significant digits provided, maximum value of $D$ reached, and approximate values of $E$ and $A$.

    \begin{table}
        \begin{center}
            \begin{small}
        {
        \ttfamily
        \begin{tabular}{lrrrrr}
            \multicolumn{1}{c}{State} & \multicolumn{1}{c}{$R$} & \multicolumn{1}{c}{$E_e$} & \multicolumn{1}{c}{$A$} & \multicolumn{1}{c}{Digits} & \multicolumn{1}{c}{$D_{\rm max}$} \\
            \hline 
            $2p\sigma_u$ & 2 &     -0.667534392202383 &     -1.186889392359195 &     98 &    100 \\
            $6p\sigma_u$ & 10 &     -0.049370966780030 &     -0.478090183465735 &    100 &    100 \\
            $5s\sigma_g$ & 10 &     -0.051428455005144 &      0.962222230928367 &    100 &     98 \\
            $6f\sigma_u$ & 8 &     -0.066255008265486 &    -10.930552412011943 &    100 &     97 \\
            $6d\pi_g $ & 10 &     -0.051519882071881 &     -4.869986869409223 &    100 &     96 \\
            $3d\pi_g $ & 4 &     -0.230953442309872 &     -5.194805350517823 &    100 &     93 \\
            $5p\pi_u $ & 10 &     -0.057271824571940 &     -1.386797316468034 &    100 &     93 \\
            $6d\sigma_g$ & 10 &     -0.060074021734383 &     -4.529352507666266 &    100 &     92 \\
            $1s\sigma_g$ & 2 &     -1.102634214494946 &      0.811729584624757 &    100 &     91 \\
            $3d\sigma_g$ & 4 &     -0.285723790479775 &     -4.860858109730897 &    100 &     90 \\
            $5d\delta_g$ & 10 &     -0.062792214839847 &     -5.531151234693738 &    100 &     90 \\
            $8h\sigma_u$ & 10 &     -0.032657740020992 &    -29.179586335030141 &     91 &     90 \\
            $5g\phi_g$ & 8 &     -0.077751893406662 &    -19.312733629824027 &    100 &     87 \\
            $5f\phi_u$ & 10 &     -0.067512161659874 &    -11.613031675139453 &    100 &     86 \\
            $6h\gamma_u$ & 10 &     -0.053894253760732 &    -29.371445454399158 &    100 &     85 \\
            $5g\gamma_g$ & 10 &     -0.071215504372313 &    -19.668697103247155 &    100 &     82 \\
            $8k\delta_u$ & 10 &     -0.031625825783903 &    -55.263865892094628 &     89 &     79 \\
            $10m\sigma_u$ & 10 &     -0.020119384615596 &    -89.495966943427064 &     80 &     71 \\
            $7i\delta_g$ & 10 &     -0.041602604901644 &    -41.056025671887276 &     80 &     64 \\
            $7i\sigma_g$ & 8 &     -0.041539060710879 &    -41.332737524441718 &     73 &     63 \\
            $9l\sigma_g$ & 10 &     -0.024922262061950 &    -71.375453234003473 &     69 &     57 \\
        \end{tabular}
}    
\end{small}
\end{center}
        \caption{Benchmark states computed in this work. The full numbers with the indicated number of significant digits are provided in the Supplementary Material. \label{tab:benchmarks}}
    \end{table}

    \subsection*{Bound states and the location of their minima}

\begin{figure}
        \begin{center}
            \includegraphics[width=\textwidth]{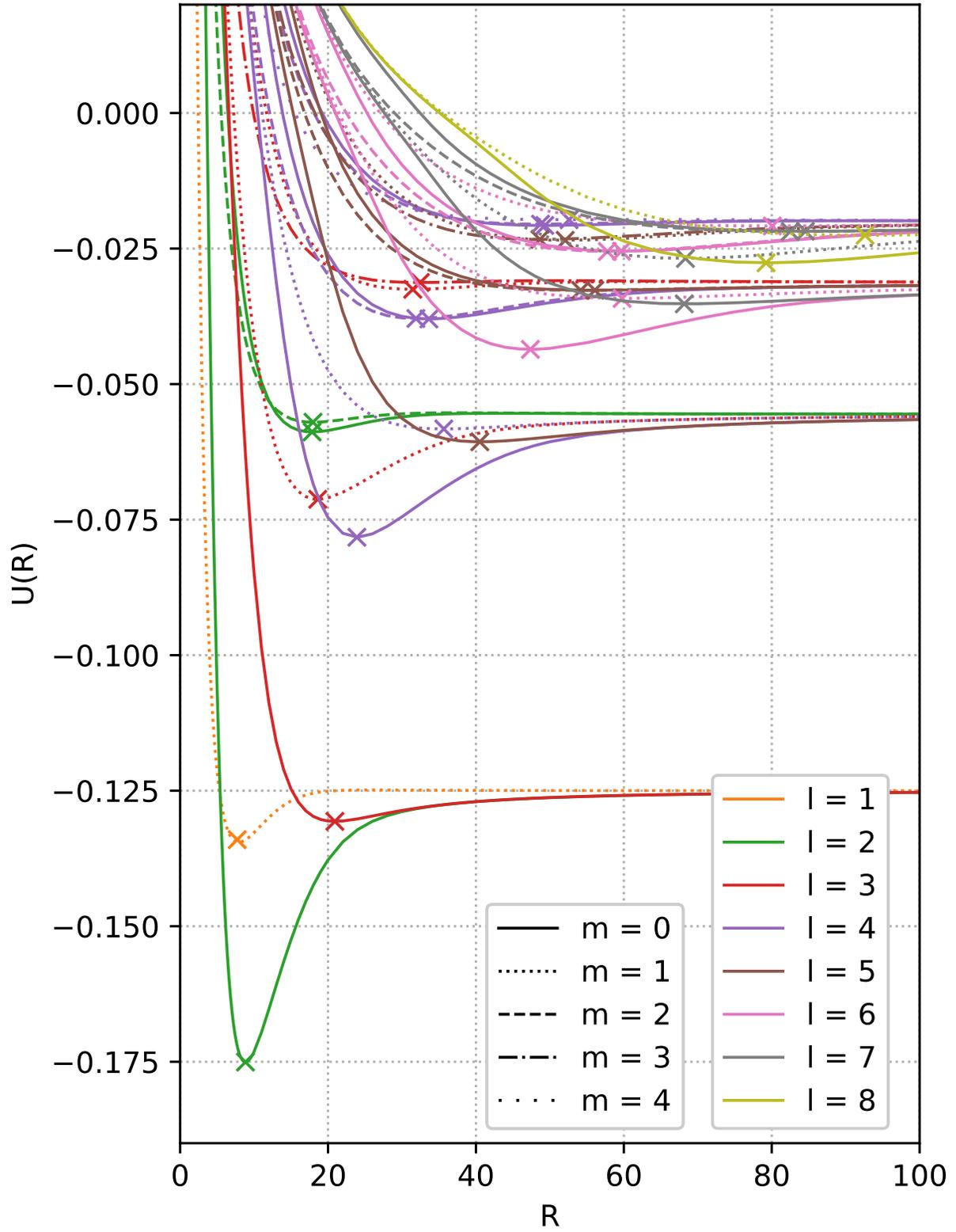} \hspace{0.5cm}
            \caption{\label{fig:bss} All the bound states computed in the present work, excluding the ground state. The equilibrium distance for each state is marked with an x.}
        \end{center}
    \end{figure}

    Within the Born-Oppenheimer approximation, a minimum of a potential-energy curve is a necessary condition for bound states.
    \begin{equation}
        U(R) = E(R) + \frac{1}{R}.
    \end{equation}
    By inspecting several of the plots of $U(R)$ for different states, it is easy to realize that several of them exhibit a minimum.
    Fig. \ref{fig:bss} shows $U(R)$ for all the binding states computed in the present work, with the exception of the ground state.
    As stated in Ref. \cite{Fernandez_1995}, equilibrium distances, energies, and separation constants, can be computed by means of the RPM by solving the additional equation
    \begin{equation}
        \frac{\partial F}{\partial A} \frac{\partial G}{\partial R} - \frac{\partial F}{\partial R}\frac{\partial G}{\partial A} = 0,
        \label{eq:minimum}
    \end{equation}
    where $F(U, A, R) = H_{\mu, D}^d(U, A, R)$, and $G(U, A, R) = H_{\lambda, D}^d(U, A, R)$.

    Therefore, to obtain the equilibrium energy $U_{\rm eq}$, separation constant $A_{\rm eq}$, and internuclear distance $R_{\rm eq}$, we solve Eqs. \eqref{eq:hankdet_roots} and \eqref{eq:minimum} simultaneously.
    For the ground state, we found that it is better to solve instead for $H_{\lambda, 2D}^d(U, A, R)$, since, for this particular set of parameters, the roots of the Hankel determinants converge faster towards the solution of Eq. \eqref{eq:mu} than Eq. \eqref{eq:lambda}.
    Analyzing those solutions with $D \leq 100$ for $d = 0, 1$, and $2$, we get the following results for the ground state:
    \begin{small}
        \begin{equation}
            \begin{aligned}
                U_{\rm eq} = &-0.6026346191065398787275621562899479553992346953448354728\\
                             &77071864391547692204240182928548052208107736708904195627167\\
                             &542817913729056948087124900979582036210907045942873,\\
                A_{\rm eq} = &0.8097945123220959277383940439312982739965337543254855548957\\
                             &26033206922628298959352111245158077673262223968225599542440\\
                             &9412145709954470705258139785977372209240315698,\\
                R_{\rm eq} = &1.9971933199699921200682981412764698139402981873092336045912\\
                             &15197873160737510275851945297613902218158798556730647200620\\
                             &903944890612509331375201735299111630413056993.\\
            \end{aligned}
        \end{equation}
    \end{small}

    We performed similar computations with $D \leq 50$ for the other bound states, and tabulated the results with 10 significant digits in Table \ref{tab:bss}. 
    The same results are provided with 40 significant digits in the Supplementary Material.
    \begin{table}
		\begin{footnotesize}
            \ttfamily
                \begin{center}
                    \begin{tabular}{llll}
                        \textrm{State} & \multicolumn{1}{c}{$R$} & \multicolumn{1}{c}{$U$} & \multicolumn{1}{c}{$A$} \\
                        \hline
                        $1s\sigma_g$ & \phantom{-}1.997193320[0] & -6.026346191[-1]     & \phantom{-}8.097945123[-1] \\
                        $2p\pi_u$ & \phantom{-}7.930714973[0] & -1.345138166[-1]     & \phantom{-}2.069815258[-2] \\
                        $3d\sigma_g$ & \phantom{-}8.834164503[0] & -1.750490359[-1]     & -1.564171919[0] \\
                        $4d\sigma_g$ & \phantom{-}1.784921705[+1] & -5.882062666[-2]     & \phantom{-}4.217727831[-1] \\
                        $3d\delta_g$ & \phantom{-}1.796959858[+1] & -5.703350664[-2]     & -2.472110245[0] \\
                        $4f\sigma_u$ & \phantom{-}2.092104113[+1] & -1.306550866[-1]     & \phantom{-}7.116425073[0] \\
                        $4f\pi_u$ & \phantom{-}1.860780308[+1] & -7.124680574[-2]     & -3.676755591[0] \\
                        $5f\pi_u$ & \phantom{-}3.145525562[+1] & -3.250735500[-2]     & -9.612135220[-1] \\
                        $4f\phi_u$ & \phantom{-}3.247412486[+1] & -3.125685627[-2]     & -6.579295566[0] \\
                        $5g\sigma_g$ & \phantom{-}2.390026713[+1] & -7.824535362[-2]     & -5.361351964[0] \\
                        $6g\sigma_g$ & \phantom{-}1.784921705[+1] & -5.882062666[-2]     & \phantom{-}4.217727831[-1] \\
                        $7g\sigma_g$ & \phantom{-}4.930661152[+1] & -2.073678351[-2]     & -9.820801505[-1] \\
                        $5g\pi_g$ & \phantom{-}3.565684224[+1] & -5.826796664[-2]     & \phantom{-}4.741584744[0] \\
                        $5g\delta_g$ & \phantom{-}3.187986608[+1] & -3.789816631[-2]     & -7.523979396[0] \\
                        $6g\delta_g$ & \phantom{-}4.873174244[+1] & -2.049852479[-2]     & -4.156610538[0] \\
                        $5g\gamma_g$ & \phantom{-}5.259706948[+1] & -1.968258155[-2]     & -1.187068111[+1] \\
                        $6h\sigma_u$ & \phantom{-}4.052059034[+1] & -6.063995570[-2]     & \phantom{-}1.367594640[0] \\
                        $7h\sigma_u$ & \phantom{-}5.608146571[+1] & -3.267896081[-2]     & \phantom{-}5.566063982[0] \\
                        $7h\pi_u$ & \phantom{-}5.206921423[+1] & -2.348082158[-2]     & -6.403536162[0] \\
                        $6h\delta_u$ & \phantom{-}5.416040079[+1] & -3.270396067[-2]     & \phantom{-}1.275260459[0] \\
                        $6h\phi_u$ & \phantom{-}4.864109832[+1] & -2.331786680[-2]     & -1.315803003[+1] \\
                        $7i\sigma_g$ & \phantom{-}4.736111515[+1] & -4.359696188[-2]     & -1.044102614[+1] \\
                        $8i\sigma_g$ & \phantom{-}5.967581513[+1] & -2.548646137[-2]     & -9.572462260[0] \\
                        $7i\pi_g$ & \phantom{-}5.976836885[+1] & -3.420356813[-2]     & -2.353500275[0] \\
                        $8i\pi_g$ & \phantom{-}8.007299332[+1] & -2.084069028[-2]     & \phantom{-}4.351689330[0] \\
                        $7i\delta_g$ & \phantom{-}5.777185203[+1] & -2.554026774[-2]     & -1.312927871[+1] \\
                        $8k\sigma_u$ & \phantom{-}6.817053314[+1] & -3.519971544[-2]     & -5.100872001[0] \\
                        $9k\sigma_u$ & \phantom{-}8.454856343[+1] & -2.180988002[-2]     & -3.351375902[0] \\
                        $8k\pi_u$ & \phantom{-}6.832548194[+1] & -2.682656672[-2]     & -1.427116244[+1] \\
                        $8k\delta_u$ & \phantom{-}8.250740043[+1] & -2.184692793[-2]     & -7.499062644[0] \\
                        $9l\sigma_g$ & \phantom{-}7.923408151[+1] & -2.762761613[-2]     & -1.675602404[+1] \\
                        $9l\pi_g$ & \phantom{-}9.261849778[+1] & -2.255205664[-2]     & -9.869012272[0] \\
                    \end{tabular}
                \end{center}
		\end{footnotesize}

        \caption{\label{tab:bss} Equilibrium internuclear distance ($R$), energy ($U$), and separation constant ($A$) of all the bound states computed in the present work. The values between brackets indicate the power of ten by which results must be multiplied. The same table is provided with 40 significant digits in the Supplementary Material.}
    \end{table}

    \clearpage

    \section*{Conclusions}

    We have shown that the RPM is able to compute the eigenenergies and separation constants of the H$_2^+$ ion-molecule very accurately.
    The values of the electronic energy and the separation constant for selected values of the internuclear distance for 69 states are provided, as well as the equilibrium parameters for 32 bound states.
    The code used to perform these computations is also provided, and results of similar accuracy can be obtained for other eigenstates.
    The scripts used to perform the computations discussed here for the spectra different values of $R$, starting from $R = 0$, are provided as well.
    To our knowledge, this is the first time computations of such accuracy are performed for this particular problem, and therefore suggest the results presented here are used as benchmarks for testing other numerical/quantum-mechanical methods.

    \section*{Supplementary material}

    A data set collection of computational results is available in Zenodo and can be accessed via \url{https://doi.org/10.5281/zenodo.5044229}.
    The software used for the computations performed in the present work is also available in Zenodo and can be accessed via \url{https://doi.org/10.5281/zenodo.5057040}.

    \bibliographystyle{ieeetr}
    \bibliography{papers.bib}
    
\end{document}